\documentclass[journal,10pt]{IEEEtran}
\usepackage{subfigure}
\usepackage{CJK}
\usepackage{algorithm}
\usepackage{algorithmic}
\usepackage{amsmath}
\usepackage{amssymb}
\usepackage{psfrag}
\usepackage{graphicx}
\usepackage{epstopdf}
\usepackage{multirow}
\usepackage{stfloats}
\usepackage{cite}
\usepackage{color}
\usepackage{arydshln}
\usepackage{enumerate}
\usepackage{bbm}
\usepackage{bm}
\usepackage{extarrows}
\usepackage{multirow}
\usepackage[dvipsnames]{xcolor}

\usepackage{amsthm}
\theoremstyle{plain}

\usepackage{xcolor}
\usepackage{changes}
\usepackage{lipsum}

\title{Over-the-Air Aggregation for Federated Learning: Waveform  Superposition and Prototype Validation}

\author{
Huayan Guo, Yifan Zhu, Haoyu Ma, Vincent K. N. Lau, \\Kaibin Huang, Xiaofan Li, Huabin Nong, and Mingyu Zhou\\

\thanks{
This work  was supported in part by the Hong Kong Innovation and Technology Fund (ITF) project  GHP-016-18GD.
The first three authors contributed equally.
}
\thanks{H. Guo, Y. Zhu, H. Ma and V. K. N. Lau are with The Hong Kong University of Science and Technology, Hong Kong (e-mail: eeguohuayan@ust.hk, kevin.zhu@connect.ust.hk, haoyu.ma@connect.ust.hk, eeknlau@ust.hk).}
\thanks{K. Huang is with The University of Hong Kong, Hong Kong (e-mail: huangkb@eee.hku.hk).}
\thanks{X. Li is with Jinan University, China (e-mail: lixiaofan@jnu.edu.cn).}
\thanks{H. Nong is with Shenzhen Institute of Radio Testing and Technology, China (e-mail: nonghuabin@srtc.org.cn).}
\thanks{M. Zhou is with Baicells Technologies Co. Ltd., China (e-mail: 18611991947@163.com).}
}

\begin{document}

\maketitle

\begin{abstract}
In this paper, we develop an orthogonal-frequency-division-multiplexing (OFDM)-based over-the-air (OTA) aggregation solution for wireless federated learning (FL). In particular, the local gradients in massive IoT devices are modulated by an analog waveform and are then transmitted using the same wireless resources. To this end, achieving perfect waveform superposition is the key challenge, which is difficult due to the existence of frame timing offset (TO) and carrier frequency offset (CFO). In order to address these issues, we propose a two-stage waveform pre-equalization technique with a customized multiple access protocol that can estimate and then mitigate the TO and CFO for the OTA aggregation. Based on the proposed solution, we develop a hardware transceiver and application software to train a real-world FL task, which learns a deep neural network to predict the received signal strength with global positioning system information. Experiments verify that the proposed OTA aggregation solution can achieve comparable performance to offline learning procedures with high prediction accuracy.
\end{abstract}


\section{Introduction}
Wireless federated learning (FL) is an emerging technique that enables a deep neural network (DNN) to be collaboratively trained by massive Internet of Things (IoT) sensors with the coordination of an edge parameter server connecting to an access point (AP), while the privacy-preserving raw dataset is stored locally without exchange
\cite{FLgoogle,FLgoogle2,FedAVG,FLopenquestion,FLchallenge2020SPM,FLsurvey2019TIST}. In particular, the IoT sensors iteratively update their local model weights/gradients locally based on their own datasets according to a broadcast common global model, and then synchronize a new global model via weights/gradients aggregation to the parameter server. As a result, the uplink weights/gradients aggregation becomes a primary bottleneck  because massive sensors will be involved but the radio resource is limited.
Much work has been done in this area to reduce the aggregation overhead based on conventional multiple access protocols that allocate dedicated radio resource to difference sensors.
In \cite{alistarh2017qsgd, bernstein2018signsgd, wang2018spars}, compression technology is utilized to alleviate the aggregation burden on weights/gradients exchanges.
In \cite{ yang2019scheduling, tran2019federated, shi2020joint, luo2020hfel, chen2020joint,amiria2021convergence,chen2020convergence,ren2020scheduling}, user scheduling algorithms are designed to reduce the number of sensors participating in every aggregation iteration.
In essence, all these solutions deal with the trade-offs between aggregation overhead and accuracy in FL. Fixing the target learning accuracy, the aggregation overhead generally grows proportionally to the number of sensors, which usually is very large, and thus it still imposes a huge requirement on the radio resource.

Recently, a new technique named over-the-air (OTA) aggregation was proposed to address this scalability issue based on the fact that the FL aggregation only concerns the summation of the local weights/gradients instead of the individual updates.
Specifically, all sensors transmit the local updates with analog uncoded modulation in one common radio resource block
by exploiting the free aggregation property in the wireless channels.
As a result, the increase of involved sensors can be beneficial with a proper design since the total power of the aggregated signals increases.
Most existing OTA-FL investigations focus on user scheduling and power control to minimize the mean squared error (MSE) of the aggregated weights/gradients, and the modification on the learning algorithm to make sure the FL problem can converge to the same optimum as the noise-free case.
In \cite{Gunduz2020FLAriCompress,DingZ2020FLAircomp,KB2019TWCaircompFL,Amiri2019CompressAIRFLGuass}, a channel-inversion transmitter is proposed to extract an unbiased estimator of the updated weights/gradients.
In \cite{myiot,Power_control}, time-varying precoding is proposed to mitigate the noise by exploiting the non-stationarity of the gradient updates.
In \cite{LQIoTOTA}, joint channel-and-data-aware user scheduling is proposed with dynamic residual feedback to guarantee the training convergence.
In \cite{LQvertical}, a hierarchical OTA-FL framework is proposed for the vertical data partitioning.
The authors in \cite{Sery2019ConvergeAIRFL,heavy_tail} prove that the training may converge even without transmit power control and user scheduling by designing a dynamic learning rate.

Existing works on the OTA technique all suppose perfect superposition of the analog waveforms coming from different sensors since a similar idea has already been verified and applied for modulation-free remote state estimations
\cite{VinceIOTRSE,VinceTAC2019RSE1,VinceJSACmissioncriticalRSE,VinceIoT2019MAIRFusion,uncode_CEO}
and the over-the-air fusion of sensor measurements \cite{uniformforcingT,HKB2019TWCreducedimensionMIMO,HuangKB2018MIMOovertherairIOT}.
In the existing  prototype validation works \cite{demo1,demo2}, a time-domain waveform design is adopted, and the waveform misalignment is the key issue since the sensors experience different multi-path propagations and also suffer different frame timing offsets (TOs).
The  misaligned waveforms not only cause superposition distortion but also result in inter-symbol interference.
In order to alleviate this issue, the direct-sequence spread spectrum (DSSS) technique is utilized in \cite{demo1,demo2}, which multiplies each analog data symbol by a pseudo-random sequence with good correlation characteristics.
By setting a long pseudo-random sequence ($N$), a high-quality waveform superposition can be achieved, but the transmission efficiency is reduced to $1/N$.
Fortunately, this efficiency loss is negligible for remote state estimations and sensor measurements since in these scenarios the size of the data to be transmitted is small.
However, it becomes a critical challenge for OTA-FL since DNNs usually contain millions of weights, which leads to a substantial increase in the resource demand.
\begin{figure*} [!t]
    \centering
    \includegraphics[width=0.9\textwidth]{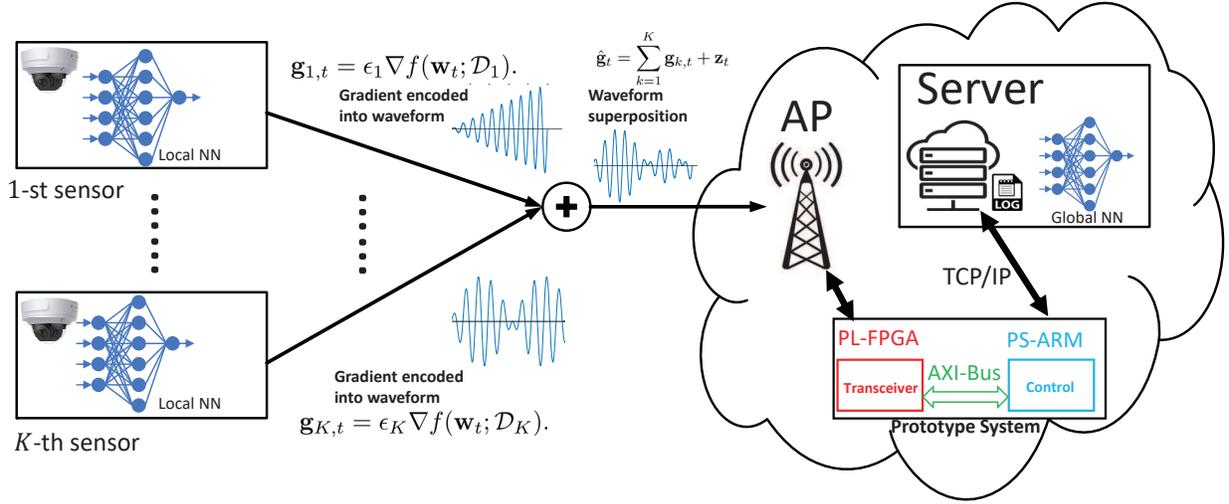}
    \caption{Illustration of the over-the-air aggregation federated learning system. Sensors train the local NN with local dataset and upload the gradients. The gradients are aggregated in a wireless channel with an analog waveform and are received by the access point (AP). The aggregated gradients are used to update the global NN in the server.}
    \label{fig:Intro_figure}
\end{figure*}

We propose an effective waveform  superposition solution with prototype
validation for the OTA-FL by utilizing the $N$-point orthogonal frequency-division multiplexing (OFDM) transmission.
In contrast to \cite{demo1,demo2}, the waveforms from different sensors can be aligned automatically in the frequency-domain sub-carriers, while the time-domain misalignment issue is addressed by the $L$-length cyclic prefixes.
Therefore, the transmission efficiency is improved to $(N-L)/N$ (while $N\gg L$).
However, a new critical issue occurs in the frequency domain.
Specifically, in OFDM transmission, the TO and the carrier frequency offset (CFO) act as frequency-domain multiplicative phase noise varying over the sub-carriers.
Since the sensors suffer different TOs and CFOs, the multiplicative phase noises are different among sensors, leading to failure of the coherent waveform superposition in the frequency domain.
To address this issue, we further propose a new protocol and algorithm to estimate and pre-compensate the phase noise with reasonable pilot overheads.
The following summarizes the critical issues to be addressed and the contributions of the paper.
\begin{itemize}
\item {\bf OFDM-based effective waveform  superposition solution for OTA-FL}:  We propose a novel waveform  superposition solution for OTA-FL by utilizing the OFDM modulation to address the waveform misalignment issue in existing works. By exploiting the circular convolution property, the waveforms from different sensors are aligned automatically in the frequency-domain sub-carriers with the aid of time-domain cyclic prefixes. The transmission efficiency can be greatly improved by setting a long OFDM symbol length.

\item {\bf Protocol and algorithm for phase error pre-compensation}:  After utilizing the OFDM, the original time-domain symbol misalignment issue is transformed to the phase noise effect in frequency domain. Inspired by broadband OFDM system imperfection analysis \cite{WLAN_Opt_TRX}, we further propose an efficient protocol to estimate and to pre-compensate the phase noise by adding one more signaling round before the consecutive OTA  frames for data transmission. The additional pilot overhead is scalable to the number of involved sensors and is also negligible compared to the huge amount of data to be transmitted.

\item {\bf Real-world prototype design and experimental validation}: 
    We develop the first real-world prototype to prove the concept of OTA-FL with the proposed waveform  superposition solution.
    The prototype is designed based on the Xilinx software defined radio (SDR) solution, in which we have one AP and two IoT sensors. A neural network (NN), which predicts the received signal strength (RSS) given global positioning system (GPS) information, is trained.
    Experimental results confirm that the OTA-FL training can be done successfully with the proposed waveform superposition solution in practice.
\end{itemize}

\section{Signal Model and Transmission Frame}
\subsection{Overview of the OTA-FL}
In FL, $K$ sensors collaboratively train a global model, which minimizes the training objective function, given as follows, with respect to the model weight parameters $\bf w$:
\begin{equation*}
\begin{aligned}[b]
{\mathcal{P}}{\text{(A)}}\quad \min_{ {\bf w}} \; &  \sum_{k=1}^K \epsilon_k f ({\bf w}; {\cal D}_k),
\end{aligned}
\end{equation*}
where $f(\cdot)$ denotes the learning model function, ${\cal D}_k$ denotes the $k$-th user's local dataset, and  $\epsilon_k$ denotes the ratio of the dataset size to the global dataset.
In the $t$-th training round, the model weight ${\bf w}_t$ is broadcast from the server to all the sensors. Each sensor then calculates the local gradient: 
\begin{equation}
{\bf g}_{k,t}=\epsilon_k \nabla f ({\bf w}_t; {\cal D}_k).
\end{equation}
Instead of updating ${\bf g}_{k,t}$ via the error-free channel by allocating dedicated resource to each user, the \textcolor{red}{AP} aggregates all the gradients in a common wireless resource, as shown in Fig. \ref{fig:Intro_figure}. If perfect waveform superposition is achieved, the server may aggregate an unbiased estimation of the true gradient by utilizing the  channel-inversion transmitter:
\begin{equation}\label{equ:estmate_g_vec_2_s2}
\begin{aligned}[b] 
{\hat{\bf g}}_t
=\sum_{k=1}^K {\bf g}_{k,t}+{\bf z}_t,
\end{aligned}
\end{equation}
where ${\bf z}_t$ is the additive noise.
Then, stochastic gradient decent (SGD) is applied to update the model, which guarantees the training may converge to a first-order optimum with a proper stepsize design:
\begin{equation}\label{equ:ideal_update}
{\bf w}_{t+1}={\bf w}_t-\eta_t {\hat{\bf g}}_t,
\end{equation}
where $\eta_t$ is the stepsize.

Our key task is to effectively realize equation \eqref{equ:estmate_g_vec_2_s2} in practice. In particular, we need a symbol-level coherent superposition of the analog waveforms from the sensors, which bears the information of local gradients ${\bf g}_{k,t}$ for all $k$ and $t$.
\subsection{Signal Model}
Denote the digital baseband time-domain symbol by $x[m]$ for $m=1,2,\dots,M$.
The corresponding continuous transmitted signal $x\left(t\right) = \sum_{m=1}^M x[m] g_T\left(t-m T_s\right)$ is constructed by $\{x[m]\}$ with a pulse-shaping filter $g_T\left(t\right)$, where $T_s$ denotes the baseband sampling period. For the conventional point-to-point transmission, the received signal is given by
\begin{align}
 r(t) = x(t) * h(t)e^{j2\pi\Delta f t} + w(t),
\end{align}
where $*$ denotes the convolution operation, $w\left(t\right)\sim\mathcal{CN}\left(0,\sigma^2\right)$ is the additive white Gaussian noise, $\Delta f$ is the CFO, and $h(t)$ is the multipath channel response, which is given by
\begin{align}\label{equ:h_time}
h(t) = \sum_{p=1}^P a_p \delta(t-\tau_p),
\end{align}
where $a_p$ and $\tau_p$ is the channel gain and path delay of the $p$-th path ($p=1,2,\dots,P$), respectively.
Next, the received signal $r\left(t\right)$ is down-sampled to discrete values with a TO (denoted by $\Delta T$) due to the imperfection of the frame synchronization, yielding
\begin{align}\label{equ:rx_1}
    r[m] =& x(t) * h(t)    \delta(t-mT_s-\Delta T)e^{j2\pi\Delta f t} + w[m]\notag\\
    =& x(t) * h(t+\Delta T)   \delta(t-mT_s)e^{j2\pi\Delta f t} + w[m].
\end{align}

We further define the effective channel as follows:
\begin{align}
    \label{equ:h_effective}
    \bar{h}(t) =&h(t+\Delta T).
\end{align}
For an $N$-point OFDM system, according to the Fourier transform relationship between the delay domain and the frequency domain, the effective channel of the $n$-th sub-carrier ($\tilde{h}[n]$) is given by
\begin{align}\label{equ:h_freq}
    \tilde{h}[n] &= \mathcal{F}\left[\bar{h}\left(t\right)\right]\delta\left(f-nf_s/N\right) \notag \\
    &=e^{j2\pi nf_s\Delta T/N} \sum_{p} a_p e^{-j2\pi nf_s\tau_p/N }\notag\\
    &= e^{j2\pi nf_s\Delta T/N} \tilde{a}[n],
\end{align}
where $\mathcal{F}\left[\cdot\right]$ denotes the Fourier transform, $f_s=1/T_s$ is the baseband sampling rate, and $\tilde{a}[n]$ is the multipath channel coefficient.

In the proposed prototype, the transmission frames consist of three basic sub-frames.
\subsubsection{Sub-Frame for Frame Timing}
The frame timing (FT) sub-frame is used to identify the starting point of a frame, which is determined when a correlation peak is observed by the match filter. 
In particular, the FT sequence adopts the up-sampled time-domain differential encoded pseudo-random  BPSK sequence as follows:
\begin{align}
    x_{\text{FT}}\left[m+2\right] = x_{\text{FT}}\left[m\right] \oplus q\left[m\right],
\end{align}
for $m=1,\dots,M_{\text{FT}}$, where $\text{mod}\left(M_{\text{FT}},2\right) = 0$. Define $x_{\text{FT}}\left[1\right] = x_{\text{FT}}\left[2\right] ~= ~1$, and
$\mathbf{q} = \left\{q\left[1\right],q\left[1\right],\dots,q\left[\frac{M_{\text{FT}}}{2}\right],q\left[\frac{M_{\text{FT}}}{2}\right]\right\}\in\{-1,1\}^{\left(M_{\text{FT}}\right)\times 1}$ is the pseudo-random BPSK sequence.
\subsubsection{ Sub-Frame for CFO Estimation}
The CFO estimation sub-frame is generated by choosing one active sub-carrier whose index is $n_{\text{CFO}}$:
\begin{align}
    {x}_{\text{CFO}}\left[m\right] = \tilde{x}\left[n_{\text{CFO}}\right] e^{j2\pi m n_{\text{CFO}}/N},~m = 1,\dots,M_{\text{CFO}},
\end{align}
where $M_{\text{CFO}} > N $ is the sequence length, and $\tilde{x}\left[n_{\text{CFO}}\right]\in\mathbb{R}$ is the frequency-domain pilot symbol.

\subsubsection{OFDM sequence}
In this prototype, we adopt OFDM for both data transmission and OTA aggregation. 
Suppose that we have already achieved a successful coarse CFO compensation, while the residual CFO is given by $\Delta f_r = \Delta f - \Delta {\hat f}$, where $\Delta {\hat f}$ is the estimated CFO.
When $\kappa N\Delta f_r T_s \ll 1$ where $\kappa$ is the number of OFDM symbols in one frame, $\Delta f_r$ can be ignored. In particular, denote the symbol in the $n$-th sub-carrier by $\tilde{s}\left[n\right]$.
The received signal in the $n$-th sub-carrier is approximated by
\begin{align}\label{eqn:f_channel}
    \tilde{r}\left[n\right] =& e^{j2\pi\Delta f_r t_0}\tilde{h}\left[n\right] \tilde{x}\left[n\right] + \tilde{w}\left[n\right],
\end{align}
where $t_0$ is the time stamp of this received symbol, and $e^{j2\pi\Delta f_r t_0}$ is the difference between $t = t_0$ and $t = 0$ in the effective channel.

\subsection{Phase Noise Issue for the Proposed OFDM-based OTA Aggregation}\label{sec:phase_noise}
\begin{figure}[!t]
    \centering
    \includegraphics[width=3in]{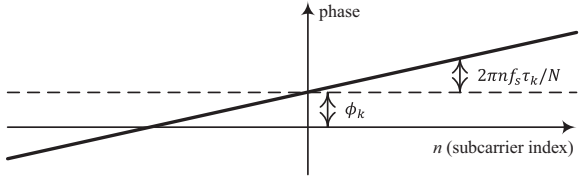}
    \caption{Phase response of the effective OTA aggregation channel of $k$-th sensor.}
    \label{fig:phase_noise}
\end{figure}

In order to realize the analog waveform superposition, a channel-inversion transmitter is suggested in \cite{Gunduz2020FLAriCompress, DingZ2020FLAircomp, KB2019TWCaircompFL, Amiri2019CompressAIRFLGuass} to pre-equalize the channel coefficients of different sensors in each sub-carrier based on the downlink estimated channel.
Combining \eqref{equ:h_freq} with the residual CFO effect, the effective downlink channel of the $k$-th sensor at time $t_{\text{DL}}^{(k)}$ is given by
\begin{align}\label{equ:effective_downlink}
    \tilde{h}_{\text{DL}}^{(k)}[n] = e^{j2\pi(\Delta f_{r}^{(k)} t_{\text{DL}}^{(k)} + nf_s\Delta T_{\text{DL}}^{(k)}/N)} \tilde{a}_k[n],
\end{align}
where $\Delta f_{r}^{(k)}$ is the residual CFO of the $k$-th sensor and $\Delta T_{\text{DL}}^{(k)}$ is the downlink TO. Note that the residual CFO of the uplink channel is opposite to that of the downlink channel. Thus, the effective uplink channel of the $k$-th sensor at time $t_{\text{UL}}^{(k)}$ is given by
\begin{align}\label{equ:effective_uplink}
    \tilde{h}_{\text{UL}}^{(k)}[n] = e^{j2\pi(-\Delta f_{r}^{(k)} t_{\text{UL}}^{(k)} + nf_s\Delta T_{\text{UL}}^{(k)}/N)} \tilde{a}_k[n],
\end{align}
where the $\Delta T_{\text{UL}}^{(k)}$ is the uplink TO. After pre-equalization, the uplink received aggregated signal of the $n$-th sub-carrier at the AP is given by
\begin{align}
\label{equ:OTA_effective}
    \tilde{r}_{\text{OTA}}[n] =& \sum_k \tilde{h}_{\text{UL}}^{(k)}[n] \frac{\tilde{x}_{k} [n]}{\tilde{h}_{\text{DL}}^{(k)}[n]}  + \tilde{w}[n] \notag \\
    =& \sum_k \underbrace{e^{j(\phi_k+2\pi nf_s\tau_k/N)}}_{\text{effective OTA channel}} \tilde{x}_{k} [n] + \tilde{w}[n],
\end{align}
where $\phi_k = -2\pi \Delta f_{r}^{(k)}(t_{\text{DL}}^{(k)}+t_{\text{UL}}^{(k)})$, and $\tau_k=\Delta T_{\text{UL}}^{(k)} - \Delta T_{\text{DL}}^{(k)}$. It is observed that the OTA waveform superposition is invalid due to the non-identical effective OTA aggregation channel, which is caused by two issues, as shown in Fig. \ref{fig:phase_noise}. First, there is an unknown phase error $\phi_k$ due to the existence of the residual CFO. Second, the imperfection of the frame timing causes a timing offset difference $\tau_k$ in OTA computation for the $k$-th sensor, which causes imperfect compensation of the channel response. Therefore, the $k$-th sensor requires $\phi_k$ and $\tau_k$ to perform a proper pre-equalization for OTA aggregation. Note that $\phi_k$ changes with $t_{\text{UL}}^{(k)}$ and $t_{\text{DL}}^{(k)}$, and $\tau_k$ may be different with a $\pm 1$ sample between two transmissions due to the synchronization offset.

\subsection{Basic Transmission Frames}
\begin{figure}[!t]
\centering
\subfigure[]{\includegraphics[width=0.3\textwidth]{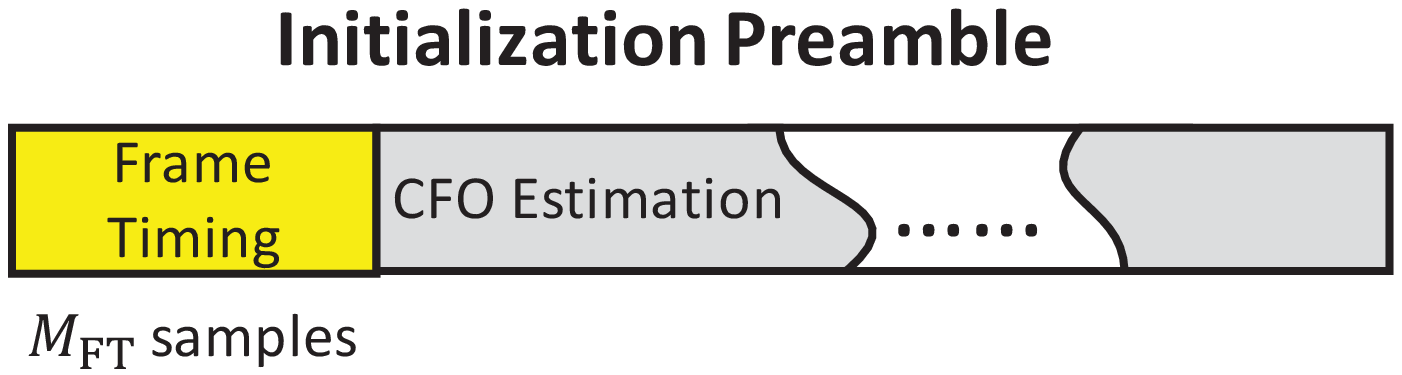}%
}
\hfil
\subfigure[]{\includegraphics[width=0.48\textwidth]{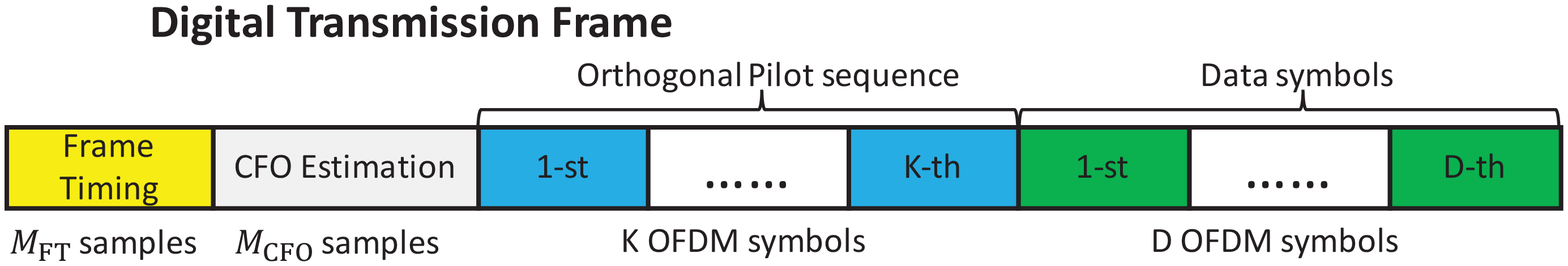}%
}
\hfil
\subfigure[]{\includegraphics[width=0.3\textwidth]{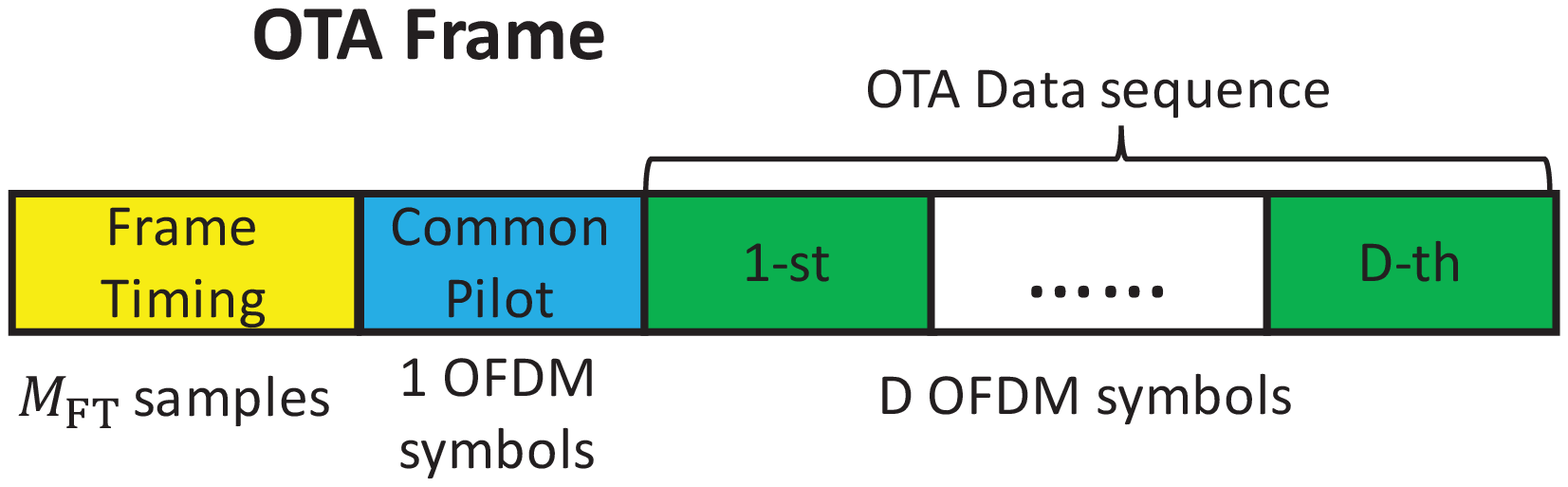}%
}
\caption{Three basic transmission frame structure. (a) is the Initialization preamble frame, (b) is the digital transmission frame, and (c) is the OTA frame.}
\label{fig:frame_str}
\end{figure}
In the proposed prototype, there are three types of transmission frame:
\begin{itemize}
    \item\textbf{Preamble for initialization:} The frame structure of the initialization preamble is shown in Fig. \ref{fig:frame_str}(a). The initialization preamble consists of two sub-frames: 1) the FT sub-frame, and 2) the CFO estimation sub-frame. In the initialization preamble we use a long CFO sequence to perform CFO estimation and adjustment at the beginning of the OTA procedure. In our prototype, the length of the CFO in the preamble is $10^6$ (i.e., 65.1 ms with a 15.36 MHz digital baseband sampling rate).
    \item\textbf{Digital transmission frame:} The frame structure of the digital transmission frame is shown in Fig. \ref{fig:frame_str}(b). Compared to the preamble for initialization, the CFO sub-frame in the digital transmission frame is shorter, which is set to the length of two OFDM symbols. In addition, we adopt an orthogonal pilot for channel estimation, which consists of $K$ OFDM symbols. The orthogonal pilot OFDM symbol contains 256 sub-carriers with a 32-length cyclic prefix, and it adopts 4-QAM modulation. Data symbols are followed with orthogonal pilots, which adopt 16-QAM.
    \item\textbf{OTA aggregation frame:} The frame structure of the OTA aggregation frame is shown in Fig. \ref{fig:frame_str}(c). In contrast to the digital transmission frame, the OTA aggregation frame does not have a CFO sub-frame. The OTA aggregation frame uses one common OFDM symbol for all sensors, and the analog data is modulated with pulse amplitude modulation in each OFDM subcarrier.
\end{itemize}

\section{Protocol and Algorithms for the Waveform Superposition}
In this section, we address the issues caused by the TO and the CFO that lead to the failure of OTA aggregation by proposing a new MAC protocol and the corresponding algorithms.
\subsection{Brief Introduction to the Proposed Protocol}
\begin{figure} [!t]
    \centering
    \includegraphics[width=0.45\textwidth]{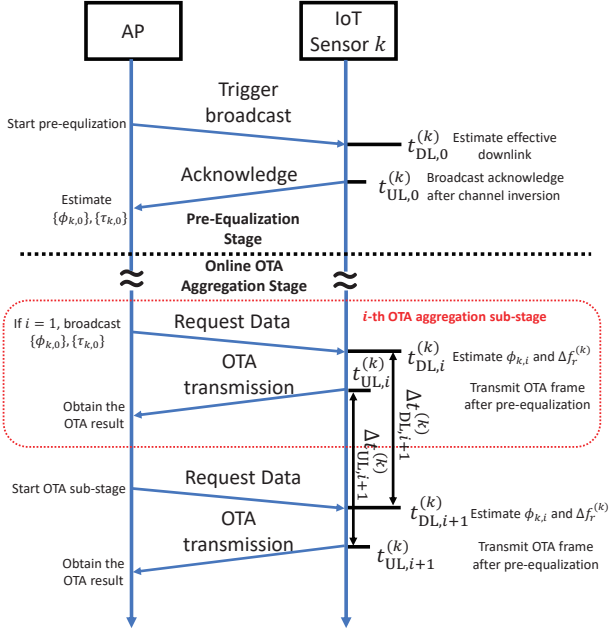}
    \caption{Illustration of the proposed physical layer OTA protocol. }
    \label{fig:protocol}
\end{figure}
In order to estimate $\phi_k$ and $\tau_k$, we proposed a physical layer protocol to jointly estimate and compensate the TO and CFO for the OTA aggregation procedure. As shown in Fig. \ref{fig:protocol}, the OTA aggregation physical layer protocol can be considered as a two-stage handshake procedure. The first stage is the pre-equlization stage, which compensates the initial CFO effect and multi-path fading. The second stage is the online OTA aggregation stage, consisting of multiple OTA aggregation sub-stages in which the instantaneous effect of the TO and residual CFO are tracked and compensated. The details of each step of the protocol are given as follows.
\begin{itemize}
    \item \textbf{Pre-equlization Stage (Downlink)}: The AP broadcasts a digital transmission frame to trigger the sensors. When the sensor $k$ receives the frame, it starts a timer and record the current time $t^{(k)}_{\text{DL},0}$. Then, the effective downlink channel $\tilde{h}_{\text{DL},0}^{(k)}$ is estimated with pilot OFDM symbols. 
    \item \textbf{Pre-equlization Stage (Uplink)}: The sensors feed back a pre-equalized digital transmission frame in \eqref{equ:OTA_effective} as acknowledgment, and records the current time $t^{(k)}_{\text{UL},0}$. With the orthogonal pilots, the AP estimates the effective OTA aggregation channel $\{\tilde{h}_{\text{OTA},0}^{(k)}\}$ in \eqref{equ:OTA_effective} to obtain the phase error $\{\phi_{k,0}\}$ and the timing offset $\{\tau_{k,0}\}$ in this stage for each sensor.
    \item \textbf{Online OTA Aggregation Stage (Downlink)}: The AP broadcasts a digital transmission frame to request the OTA aggregated data. Each sensor records the time $t^{(k)}_{\text{DL},i}$ when receiving the frame, and estimates the effective downlink channel $\tilde{h}_{\text{DL},i}^{(k)}$. Note that in the $1$-st OTA aggregation operation, $\{\phi_{k,0}\}$ and $\{\tau_{k,0}\}$ are broadcast to the sensors by the AP.
    \item \textbf{Online OTA Aggregation Stage (Uplink)}: The timer records the current time $t^{(k)}_{\text{UL},i}$. The downlink and uplink period between the previous and the current OTA aggregation sub-stage, $\Delta t_{\text{DL},i}^{(k)} = t^{(k)}_{\text{DL},i} - t^{(k)}_{\text{DL},i-1}$ and $\Delta t_{\text{UL},i}^{(k)} = t^{(k)}_{\text{UL},i} - t^{(k)}_{\text{UL},i-1}$, are calculated. Each user estimates $\phi_{k,i}$ and $\tau_{k,i}$ according to the effective downlink channel $\tilde{h}_{\text{DL},i}^{(k)}$, the downlink period $\Delta t_{\text{DL},i}^{(k)}$, and the uplink period $\Delta t_{\text{UL},i}^{(k)}$. Then the OTA aggregation frame is transmitted after pre-equalization with estimated $\phi_{k,i}$ and $\tau_{k,i}$ simultaneously, which is shown in Section \ref{sec:phase_noise_compensation}.
\end{itemize}

\subsection{Phase Error Estimation and Compensation} \label{sec:phase_noise_compensation}
In the proposed protocol, the OTA pre-equalization requires the estimated phase noise $\{\hat{\phi}_k\}$ and the estimated timing offset $\{\hat{\tau}_k\}$. Specifically, $\{\hat{\phi}_{k,0}\}$ and $\{\hat{\tau}_{k,0}\}$ are estimated in the pre-equlization stage at the AP side. As illustrated in Section \ref{sec:phase_noise}, $\hat{\phi}_k$ changes with $t_{\text{DL}}$ and $t_{\text{UL}}$, such that in the $i$-th OTA aggregation sub-stage uplink step it changes to $\phi_{k,i} = e^{-j2\pi \Delta f_{r}^{(k)}(\Delta t_{\text{DL},i}^{(k)}+\Delta t_{\text{UL},i}^{(k)})}\phi_{k,i-1}$. As a result, the residual CFO $\Delta f_{r}^{(k)}$ needs to be estimated to perform correct OTA aggregation. We first introduce the estimation method for the desired variables. The estimation of the channel $\tilde{h}$ is denoted by $\hat{h}$.
\subsubsection{Estimation of $\{\phi_{k,0}\}$ at the AP}
As shown in Fig. \ref{fig:phase_noise}, $\phi_{k,0}$ reflects on the intercept of the effective OTA aggregation channel phase response in the pre-equlization stage. Thus, it can be estimated through the average phase of the effective OTA aggregation channel as
\begin{align}
    \hat{\phi}_{k,0} = \frac{1}{N-2}\sum_{n=1}^{N/2-1}(\angle \hat{h}_{\text{OTA},0}^{(k)}[-n] + \angle \hat{h}_{\text{OTA},0}^{(k)}[n]),
\end{align}
where we restrict $\angle \hat{h}_{\text{OTA},0}^{(k)}[n+1]-\angle \hat{h}_{\text{OTA},0}^{(k)}[n] \in (-\pi,\pi]$ to avoid the phase ambiguity.
\subsubsection{Estimation of $\{\tau_{k,0}\}$ at the AP}
As shown in Fig. \ref{fig:phase_noise}, $\tau_{k,0}$ reflects on the slope of the effective OTA channel phase response in the pre-equlization stage. Thus, it can be estimated by taking the average of the phase increment of the effective OTA aggregation channel between the adjacent sub-carriers:
\begin{align}
    \hat{\tau}_{k,0} = \frac{N}{2\pi f_s} \angle \sum_{n=-N/2}^{N/2-2} \hat{h}_{\text{OTA},0}^{(k)*}[n] \cdot \hat{h}_{\text{OTA},0}^{(k)}[n+1].
\end{align}

\subsubsection{Estimation of $\Delta f_{r}^{(k)}$ at the $k$-th Sensor}
From \eqref{equ:effective_downlink}, the relationship between the two effective downlink channels $\tilde{h}_{\text{DL},i-1}^{(k)}$ and $\tilde{h}_{\text{DL},i}^{(k)}$ is given by
\begin{align}\label{equ:downlink_relation}
    \tilde{h}_{\text{DL},i}^{(k)}[n] = e^{j2\pi\Delta f_{r}^{(k)} \Delta t_{\text{DL}}^{(k)}} \tilde{h}_{\text{DL},i-1}^{(k)}[n].
\end{align}
So the residual CFO $\Delta f_{r}^{(k)}$ can be estimated by
\begin{align}
    \Delta \hat{f}_{r}^{(k)} = \frac{1}{2\pi\Delta t_{\text{DL},i}^{(k)}} \angle \sum_{n=-N/2}^{N/2-1} \hat{h}_{\text{DL},i-1}^{(k)*}[n] \cdot \hat{h}_{\text{DL},i}^{(k)}[n].
\end{align}
Note that when $|\Delta f_{r}^{(k)}\Delta t_{\text{DL},i}^{(k)}|\geq 1/2$, there will be phase ambiguity, causing the error of $\Delta \hat{f}_{r}^{(k)}$. So the residual CFO must satisfy
\begin{align}\label{equ:residual_CFO_r}
    |\Delta f_{r}^{(k)}| < \frac{1}{2\Delta t_{\text{DL},i}^{(k)}}.
\end{align}\par
\subsubsection{Phase Noise Compensation in the OTA Aggregation Stage}
With the above estimation results, the pre-equalization in the OTA aggregation stage can be designed. First, $\hat{\phi}_{k,i}$ is given by
\begin{align}
    \hat{\phi}_{k,i} = e^{-j2\pi \Delta \hat{f}_{r}^{(k)}(\Delta t_{\text{DL},i}^{(k)}+\Delta t_{\text{UL},i}^{(k)})}\hat{\phi}_{k,i-1}.
\end{align}
Then, the $\pm 1$ sample difference between $\hat{\tau}_{k,i}$ and $\hat{\tau}_{k,i-1}$ can be obtained by comparing  $\tilde{h}_{\text{DL},i}^{(k)}$ and $\tilde{h}_{\text{DL},i-1}^{(k)}$.
Finally, the pre-equalized symbol at sensor $k$ in the $i$-th OTA aggregation operation is given by
\begin{align}\label{equ:equ_stage_2}
    \tilde{x}_{k,e}[n] = \frac{1}{e^{j(\hat{\phi}_{k,i}+2\pi nf_s \hat{\tau}_{k,i}/N)} \hat{h}_{\text{DL},i}^{(k)}[n]} \tilde{x}_{k}[n],
\end{align}
Combining \eqref{equ:OTA_effective}, \eqref{equ:downlink_relation} and \eqref{equ:equ_stage_2}, the aggregated signal at the AP is given by
\begin{align}
    \tilde{r}_{\text{OTA}}[n] =& \sum_k \tilde{h}_{\text{UL},i}^{(k)}[n] \tilde{x}_{k,e} [n] + \tilde{w}[n] \notag \\
    =& \sum_k \tilde{x}_{k} [n] + \tilde{w}[n],
\end{align}
which satisfies the requirements of waveform superposition for the OTA aggregation procedure.

\subsection{Carrier Frequency Offset Estimation} \label{sec:CFO_estimation}
Effectiveness of the proposed protocol for waveform superposition requires residual CFO $|\Delta f_{r}^{(k)}| < 1/(2\Delta t_{\text{DL},i}^{(k)})$. Thus, CFO compensation and residual CFO tracking is critical. To address this issue, we propose a two-step method to compensate the CFO for the OTA aggregation procedure: 1) initialization of coarse CFO correction, and 2) residual CFO tracking. Specifically, initialization of CFO correction is performed immediately after the prototype enters the OTA aggregation state, and residual CFO tracking is performed to monitor the value of the residual CFO.\par
\subsubsection{Coarse CFO Estimation} At the system startup stage, an initialization of CFO correction is performed to guarantee the residual CFO is within the requirement of the system. Specifically, the CFO estimation is performed with the initialization preamble with a maximum likelihood estimator. Given the received preamble $r_{\text{Init}}\left[m\right]$, the estimated coarse CFO $\Delta \hat{f}_c$ is given by \cite{coarse_CFO}
\begin{align}
    \Delta\hat{f}_{c} = \frac{1}{2\pi T_s L_{\text{SPAN}}}\angle \sum_{m = 1}^{M_{\text{CFO}}}r_{\text{Init}}^*\left[m\right]r_{\text{Init}}\left[m+L_{\text{SPAN}}\right],
\end{align}
where $T_s$ is the baseband sampling period, and $L_{\text{SPAN}}$ is the length of the span, which satisfies that $\angle r^{*}_{\text{Init}}\left[m+L_{\text{SPAN}}\right]{r}_{\text{Init}}\left[m\right] = 0$. As such, the coarse CFO estimation range is $\Delta \hat{f}_c \in \left(-\frac{1}{2T_s L_{\text{SPAN}}},\frac{1}{2T_s L_{\text{SPAN}}}\right]$.\par

\subsubsection{CFO Tracking} The residual CFO is tracked to ensure that it satisfies the requirement of the OTA aggregation protocol since the CFO usually varies slowly. Given the received signal with residual CFO $\Delta f_r$ by (13), and that the channel is unchanged during the OTA aggregation procedure, the received signal in the $n$-th subcarrier at time $\text{t}_p$ is denoted as $\tilde{r}\left[n\right]\left(\text{t}_p\right)$, which can be obtained by
\begin{align}
    \tilde{r}\left[n\right]\left(\text{t}_p\right) =& e^{j2\pi\Delta f_r \text{t}_p}\tilde{h}\left[n\right] \tilde{x}\left[n\right] + \tilde{w}\left[n\right].
\end{align}
The estimated residual CFO $\Delta \hat{f}_r$ can be obtained by periodically sending an identical pilot frame. Specifically, for $N$-point OFDM symbols, the $\Delta \hat{f}_r$ at time $\text{t}_p$ is given by
\begin{align}
    \Delta \hat{f}_r\left(\text{t}_p\right) &= \frac{1}{\left(\text{t}_p - \text{t}_{p-1}\right)N}\sum_{n=1}^N\angle \frac{\tilde{r}\left[n\right]\left(\text{t}_p\right)}{\tilde{r}\left[n\right]\left(\text{t}_{p-1}\right)}\notag\\
    &= \Delta f_r + \phi_w,~p = 1,\dots,P,
\end{align}
where $\phi_w\sim\mathcal{N}\left(0,\sigma^2_{\phi_w}\right)$ is the phase noise caused by additive Gaussian noise. Given (14), the sequential estimation of the CFO is denoted as $\Delta \bar{f}_r\left(\text{t}_p\right)$, which is given by
\begin{align}
    \Delta \bar{f}_r \left(\text{t}_p\right) = \frac{p-1}{p}\Delta \bar{f}_r\left(\text{t}_{p-1}\right) + \frac{1}{p}\Delta \hat{f}_r\left(\text{t}_p\right),~\forall p.
\end{align}

\section{Developing Key Modules}
\begin{figure} [!t]
    \centering
    \includegraphics[width=0.45\textwidth]{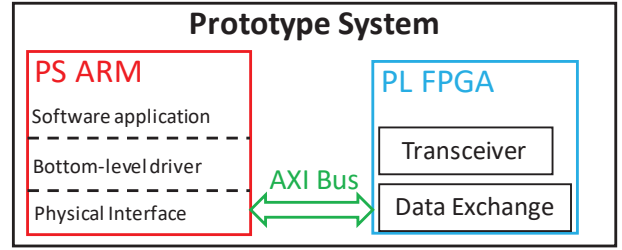}
    \caption{Structure of the prototype system. There are two parts in the system: 1) the PS ARM part, which runs the software application and driver, and 2) the PL ARM part, which runs the physical layer transceiver modules. The data exchange between PS-PL is handled by an AXI bus.}
    \label{fig:proto_system}
\end{figure}
As shown in Fig. \ref{fig:proto_system}, the prototype system consists of two major parts: 1) the programmable software (PS) ARM, and 2) the programmable logic (PL) FPGA. The key signal processing modules are designed in PL, and these modules are controlled by PS. The data exchange between PS-PL is performed with a physical AXI bus. 
\subsection{Control Logic Designs in the ARM Platform}

\begin{figure} [!t]
    \centering
    \includegraphics[width=0.45\textwidth]{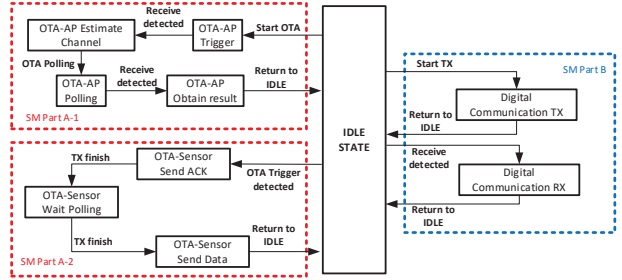}
    \caption{The state machine (SM) for control logic in the prototype. The system initialization state is the IDLE state, and the state transition depends on last state and input signals. The SM is divided into two parts: part A is for the OTA and part B is for digital communication.}
    \label{fig:SM}
\end{figure}
The control logic for the transceiver is implemented as a state machine (SM), which is shown in Fig. \ref{fig:SM}. As the figure shows, each block represents a state, and the state transition depends on the previous state and current input signals. The SM can be divided into two parts. In part A, the SM provides the control logic for OTA aggregation operation, and in part B, it provides the control logic for digital communication. The SM part A is implemented according to the proposed protocol.
\subsection{Transceiver Module Designs in the FPGA}
\begin{figure} [!t]
    \centering
    \includegraphics[width=0.45\textwidth]{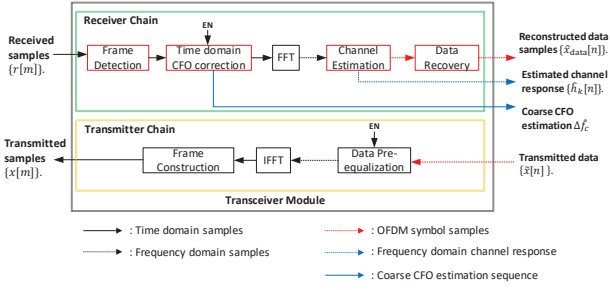}
    \caption{Physical layer transceiver structure. The upper link is the receiver channel and the bottom link is the transmitter channel. Two enable signals are controlled by the SM control logic for either digital communication purposes or OTA purposes.}
    \label{fig:TRX_Structure}
\end{figure}
The proposed prototype performs real-time signal processing in the PL FPGA with a customized transceiver module. The structure of the transceiver module is shown in Fig. \ref{fig:TRX_Structure}. As shown in the figure, the transceiver consists of a transmitter chain and a receiver chain consisting of four key sub-modules: 1) the frame detection and timing module, 2) the CFO estimation and compensation module, 3) the channel estimation and compensation module, and 4) the data pre-equalization module. 
\subsubsection{Frame Detection and Timing}
\begin{figure}[!t]
\centering
\subfigure[]{\includegraphics[width=0.24\textwidth]{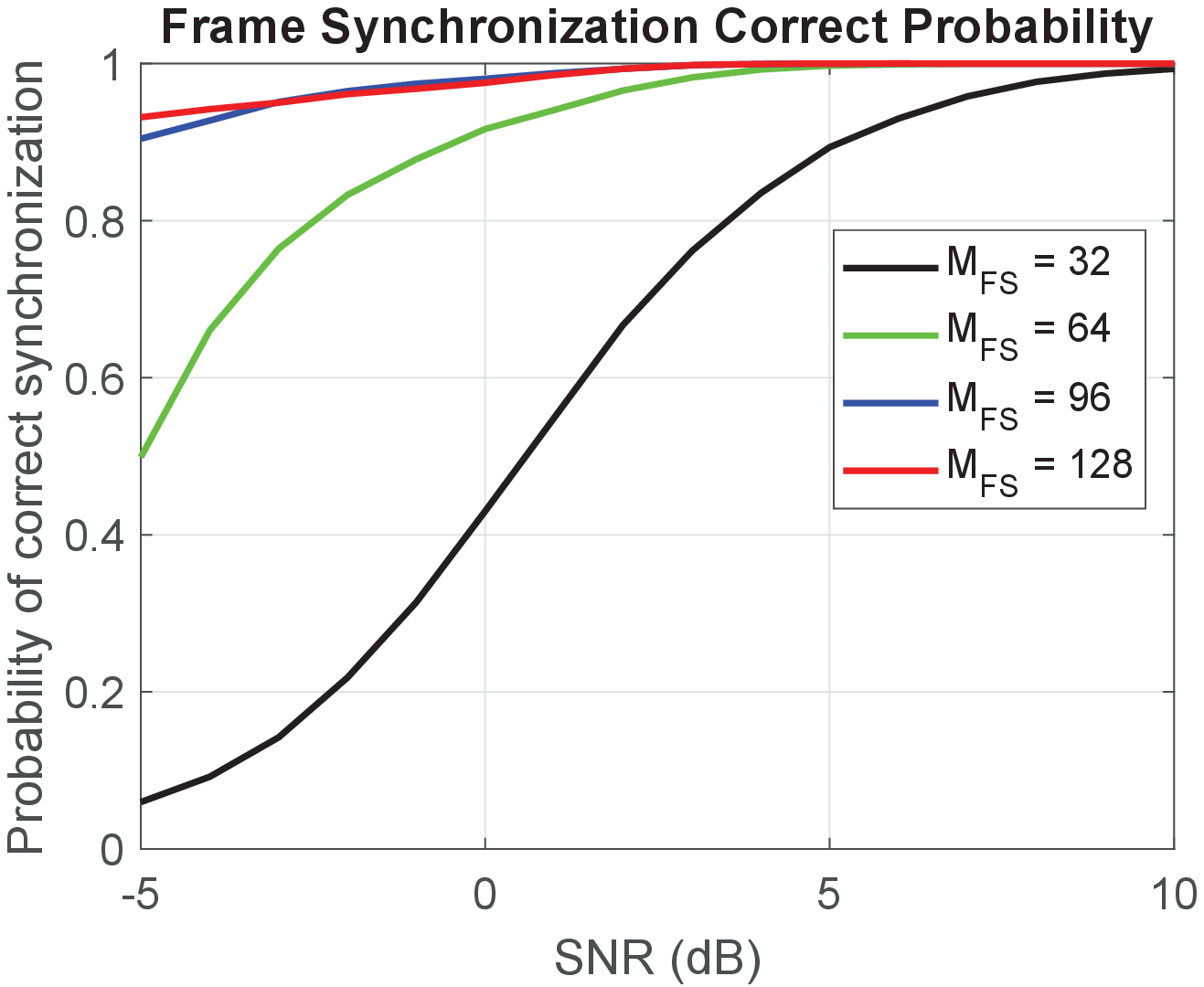}%
}
\hfil
\subfigure[]{\includegraphics[width=0.24\textwidth]{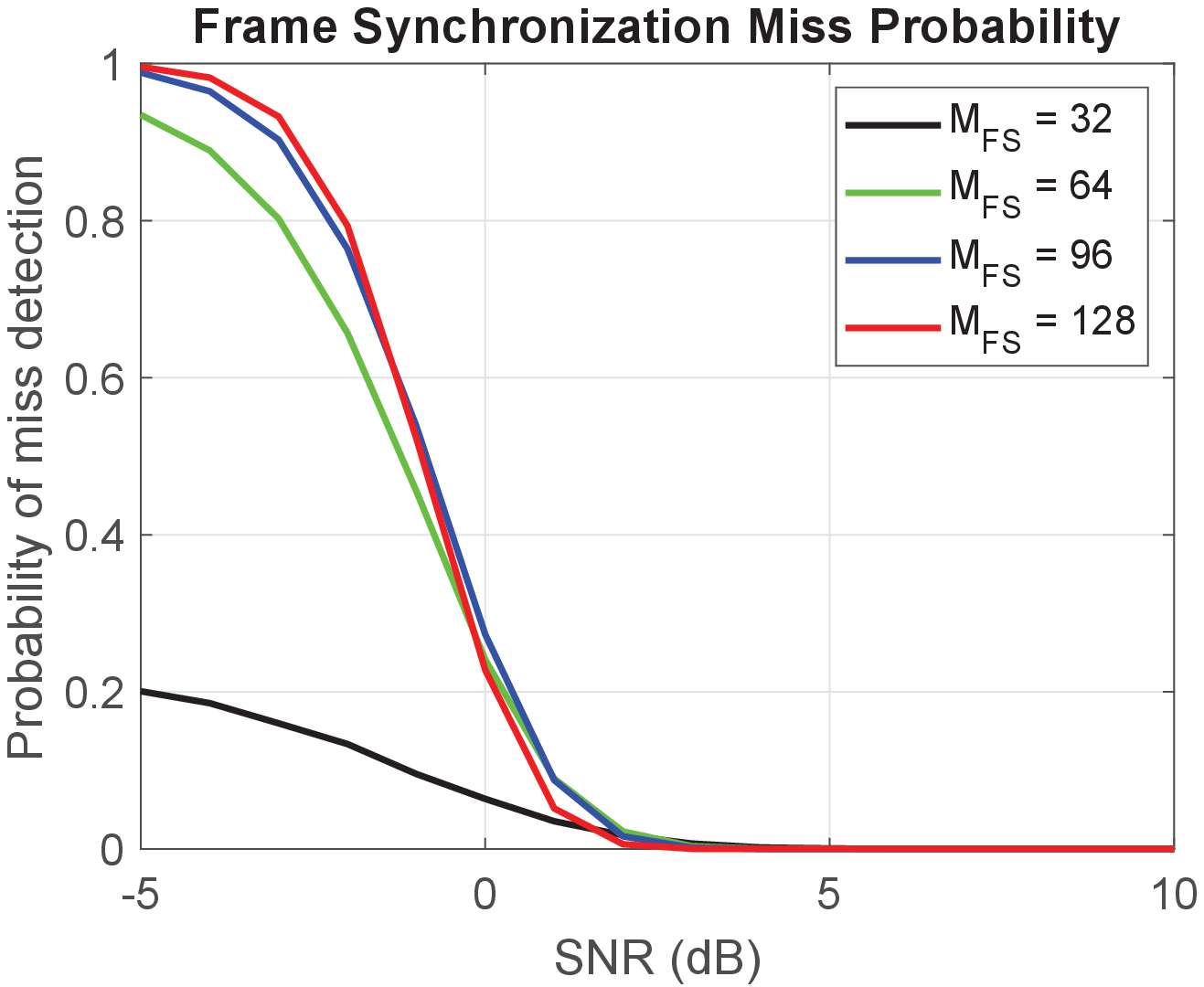}%
}
\caption{Performance of frame timing. (a) The probability of correct synchronization when a valid correlation peak is detected, and (b) The probability of the correlation peak detection.}
\label{fig:FS_Performance}
\end{figure}
The frame detection and timing module finds the start point of a received frame, which is realized with a differential decoder and correlator. To be specific, the received samples are first differentially decoded, and then cross correlation is performed with the known PR BPSK sequence as
\begin{align}
    \hat{q}\left[m\right] &= \text{sign}\left(r\left[m\right]r^*\left[m+2\right]\right),\\
    \text{Corr}\left[m'\right] &= \sum_{m = -\infty}^{\infty}\hat{q}\left[m\right]q\left[m-m'\right],
\end{align}
where $\text{Corr}\left[m'\right]$ is the output result of the correlation sequence. Under a noise-free channel, the maximum correlation output value is equal to $M_{\text{FT}}-2$. The start point of a frame $m_0$ is obtained as
\begin{align}
    m_0= \text{arg}\max_{m'} \left|\text{Corr}\left[m'\right]\right| - M_{\text{FT}} + 1.
\end{align}
A threshold $\gamma_{th}$ is defined so that only $\text{Corr}\left[m_0\right] \geq \gamma_{th}$ is denoted as a valid correlation output, and we set $\gamma_{th} = \frac{M_{\text{FT}}-2}{2}$. As shown in Fig. \ref{fig:FS_Performance}, a longer FT sequence achieves a higher probability of correct frame timing and a higher probability of miss detection.
\subsubsection{Carrier Frequency Offset Estimation and Correction}
\begin{figure}[!t]
\centering
\subfigure[]{\includegraphics[width=0.24\textwidth]{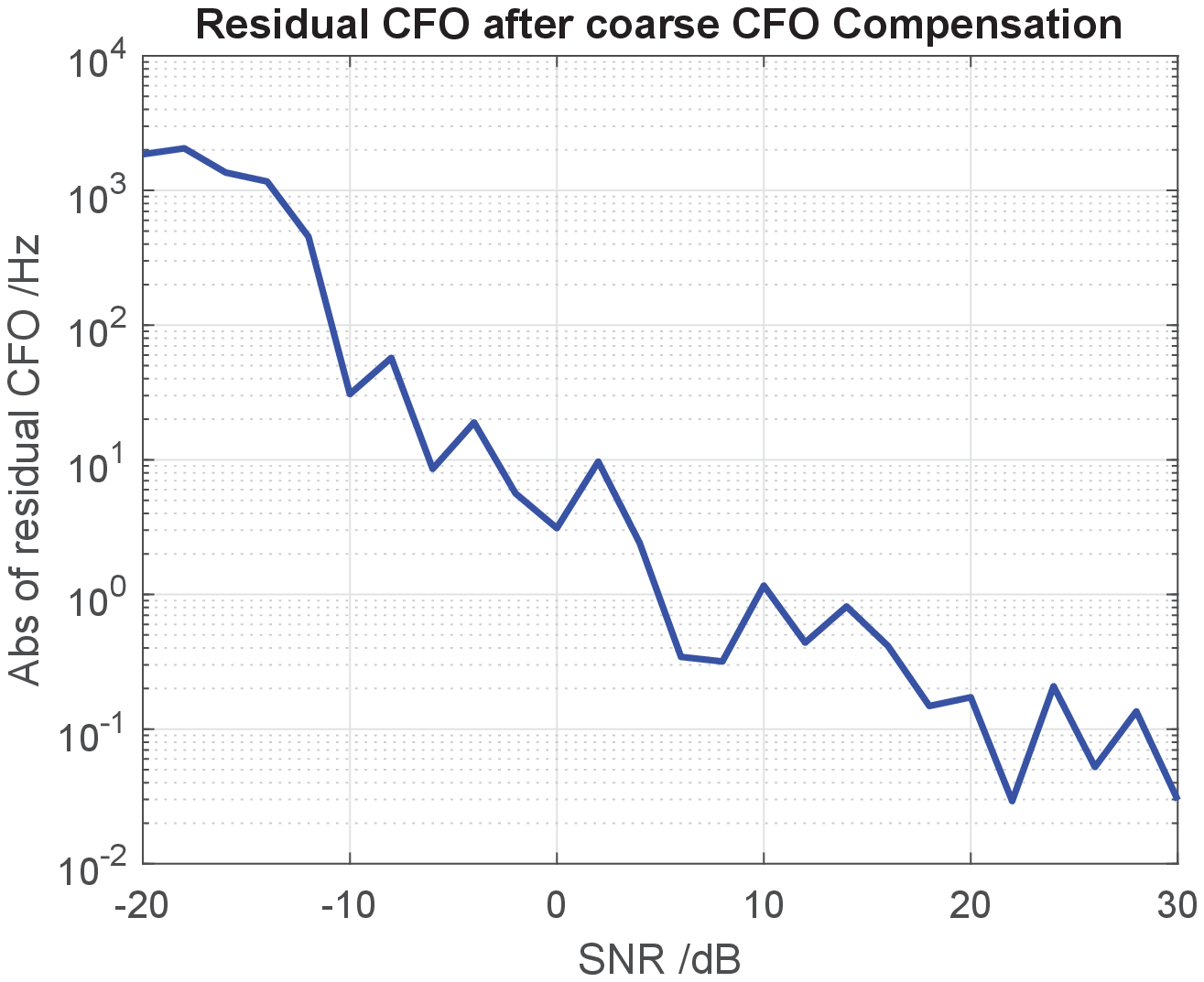}%
}
\hfil
\subfigure[]{\includegraphics[width=0.24\textwidth]{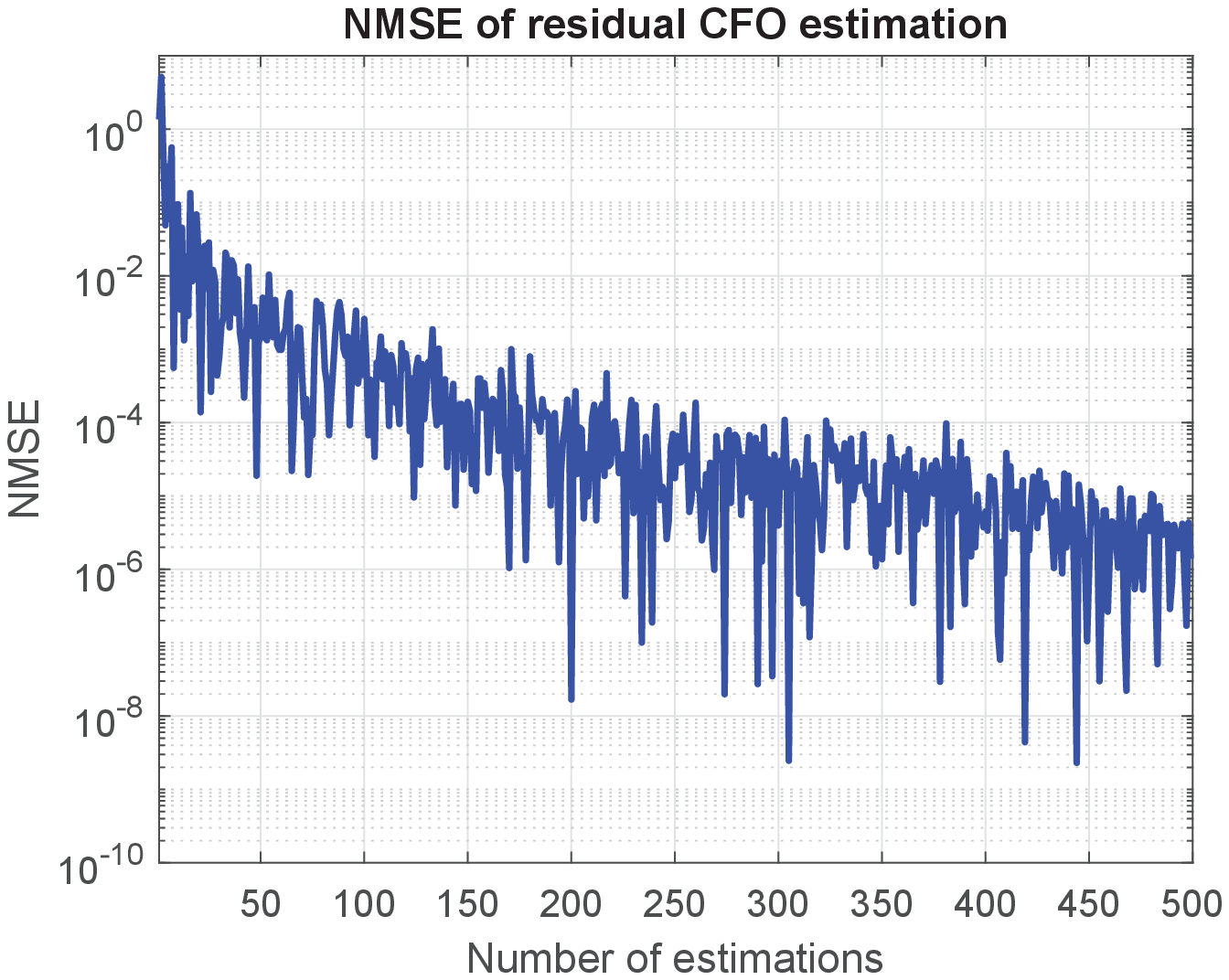}%
}
\caption{The CFO estimation performance. Coarse CFO estimation sequence length is $10^6$. (a) The residual CFO after coarse CFO estimation and compensation, and (b) The NMSE of residual CFO estimation with SNR = 0 dB. }
\label{fig:CFO_Estimation}
\end{figure}
After the start point of the received frame, the coarse CFO estimation sequence can be obtained. As introduced in Section \ref{sec:CFO_estimation}, the coarse CFO estimation is performed with a maximum likelihood estimator. The estimated coarse CFO is compensated for both digital transmission and the OTA procedure. To evaluate the performance of the coarse CFO estimation, we define the normalized mean square error (NMSE) of the estimated coarse CFO as
\begin{align}
    \text{NMSE}_{\text{CFO}} = \frac{\|\Delta \hat{f}_c - \Delta f_c\|^2_2}{\|\Delta f_c\|^2_2}.
\end{align}
As shown in Fig. \ref{fig:CFO_Estimation}(a) the residual CFO after coarse CFO estimation and compensation decreases with an increasing SNR. With a SNR over 0 dB, the residual CFO after coarse CFO compensation is within 10 Hz, while the requirement of maximum residual CFO in our prototype is 500 Hz.

After the coarse CFO is estimated and compensated, the residual CFO is estiamted and tracked sequentially by using the algorithm in Section \ref{sec:CFO_estimation}. As shown in Fig. \ref{fig:CFO_Estimation}(b), the NMSE of $\Delta \bar{f}_r \left(\text{t}_p\right)$ is decreasing with respect to the number of estimations. Denoting the sequential estimation of the residual CFO at time stamp $t_p$ as $\Delta \bar{f}_r \left(\text{t}_p\right)$, the estimation period is given as $\Delta \text{t} = \text{t}_p - \text{t}_{p-1},\forall p$. The system will request a new coarse CFO correction if $\Delta \text{t} \Delta \bar{f}_r > 1/2$.

\subsubsection{Channel Estimation and OFDM Symbol Reconstruction}
\begin{figure}[!t]
\centering
\subfigure[]{\includegraphics[width=0.24\textwidth]{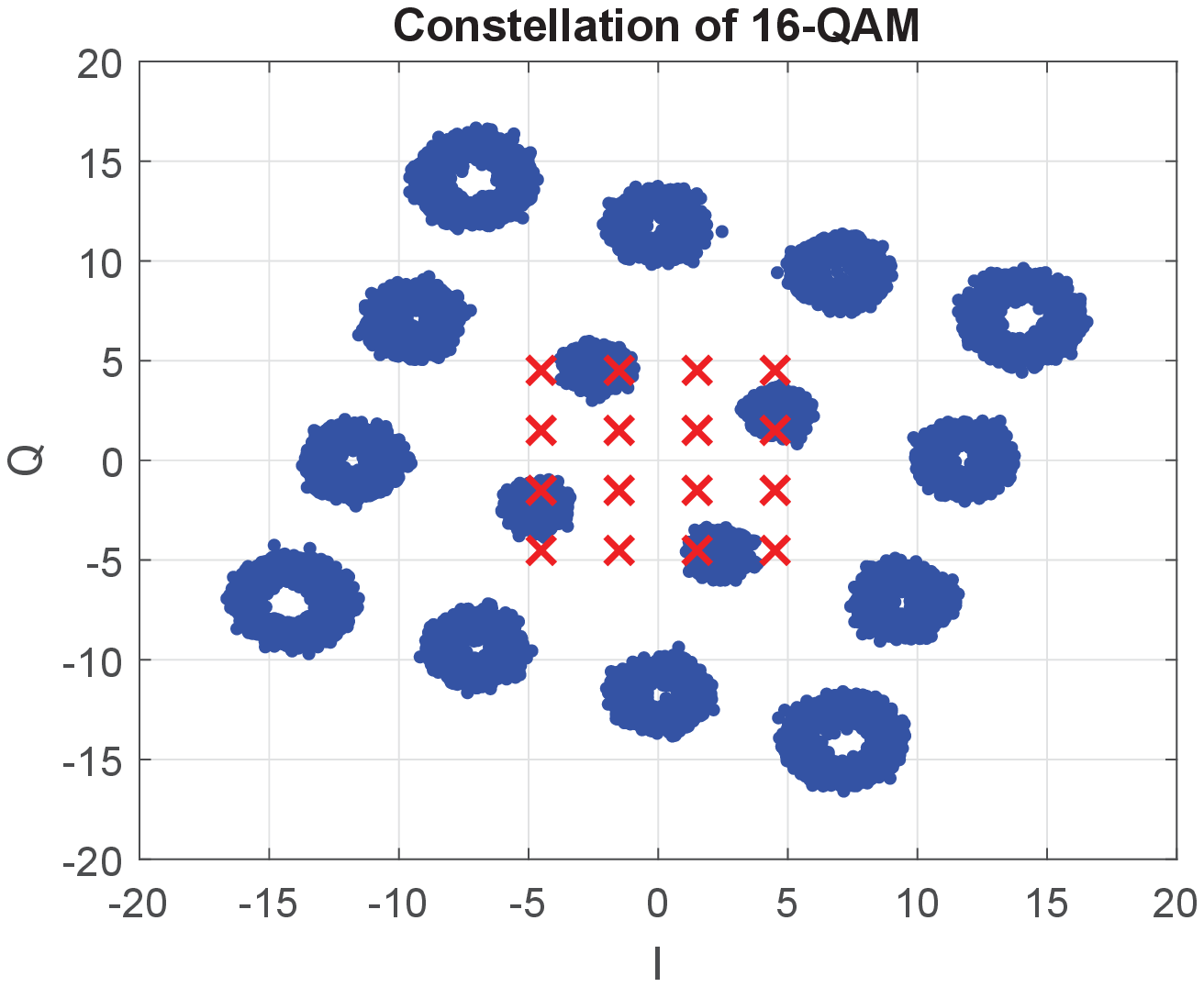}%
}
\hfil
\subfigure[]{\includegraphics[width=0.24\textwidth]{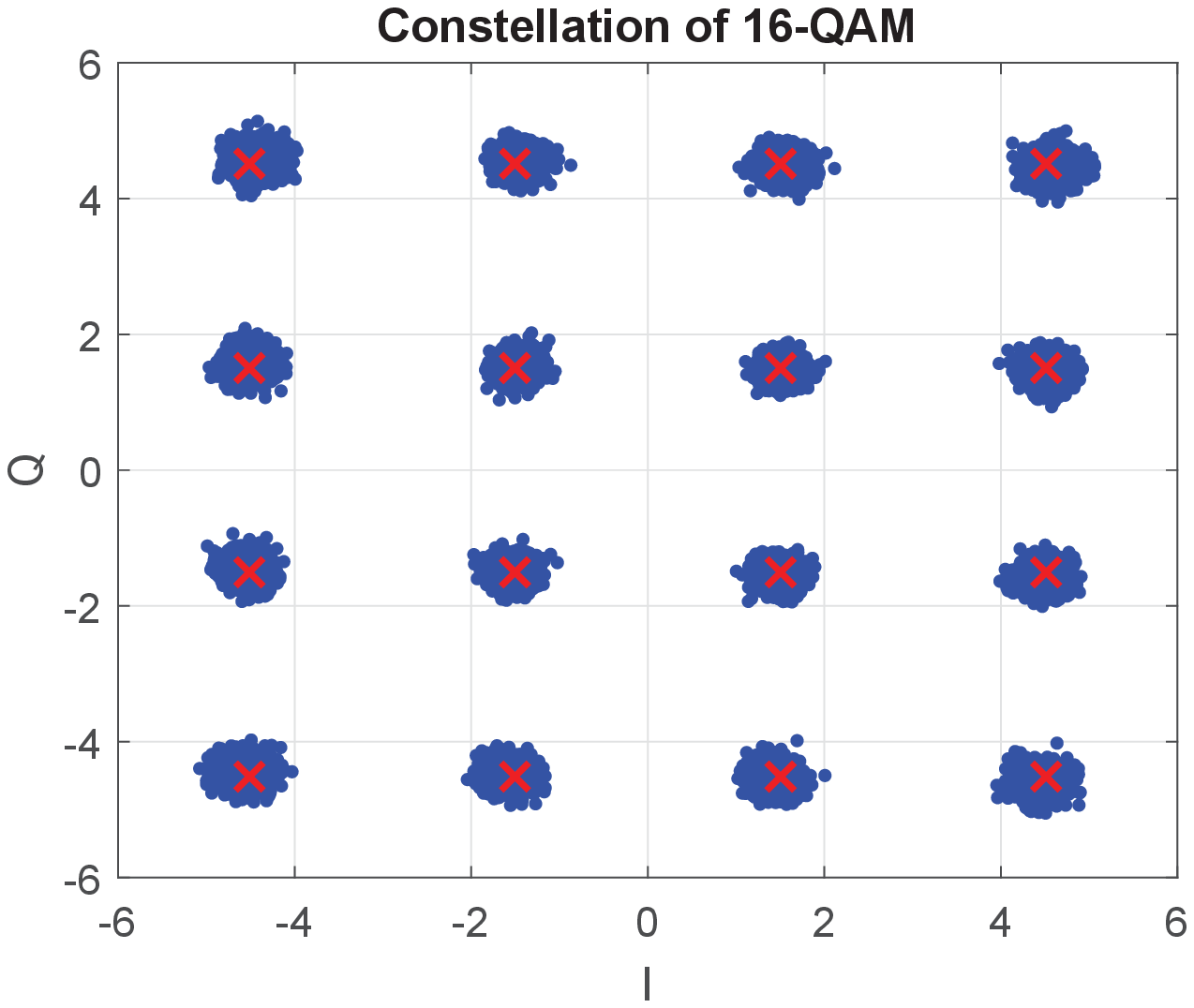}%
}
\caption{Constellation of the received OFDM symbols before and after channel equalization. The received SNR is 30 dB. (a) The channel response is not compensated, and (b) The channel response is estimated and compensated.}
\label{fig:Constellation}
\end{figure}
The channel estimation is performed with a least-square estimator in each sub-carrier. The received pilot OFDM symbols are obtained with DFT as $\tilde{\mathbf{r}}_{\text{pilot}}^{'} = \text{DFT}\left(\mathbf{r}_{\text{pilot}}^{'}\right)$. Thus, the estimated channel coefficients in the $n$-th sub-carrier $\hat{h}\left[n\right]$ is given by
\begin{align}
    \hat{h}\left[n\right] =  \frac{\tilde{r}_{\text{pilot}}^{'}\left[n\right]}{\tilde{x}_{\text{pilot}}\left[n\right]},~n = 1,\dots,N.
\end{align}
The OFDM symbols $\hat{\mathbf{x}}_{\text{data}}$ are reconstructed with estimated channel coefficients as $\hat{x}_{\text{data}}\left[n\right] = \hat{h}^{-1}\left[n\right]\tilde{r}_{\text{data}}^{'}\left[n\right],~n=1,\dots ,N$. As shown in Fig. \ref{fig:Constellation}, after the channel estimation and compensation, the received 16-QAM OFDM symbols can be decoded correctly.\par
\subsubsection{Data pre-equalization module}
\begin{figure}[!t]
\centering
\subfigure[]{\includegraphics[width=0.24\textwidth]{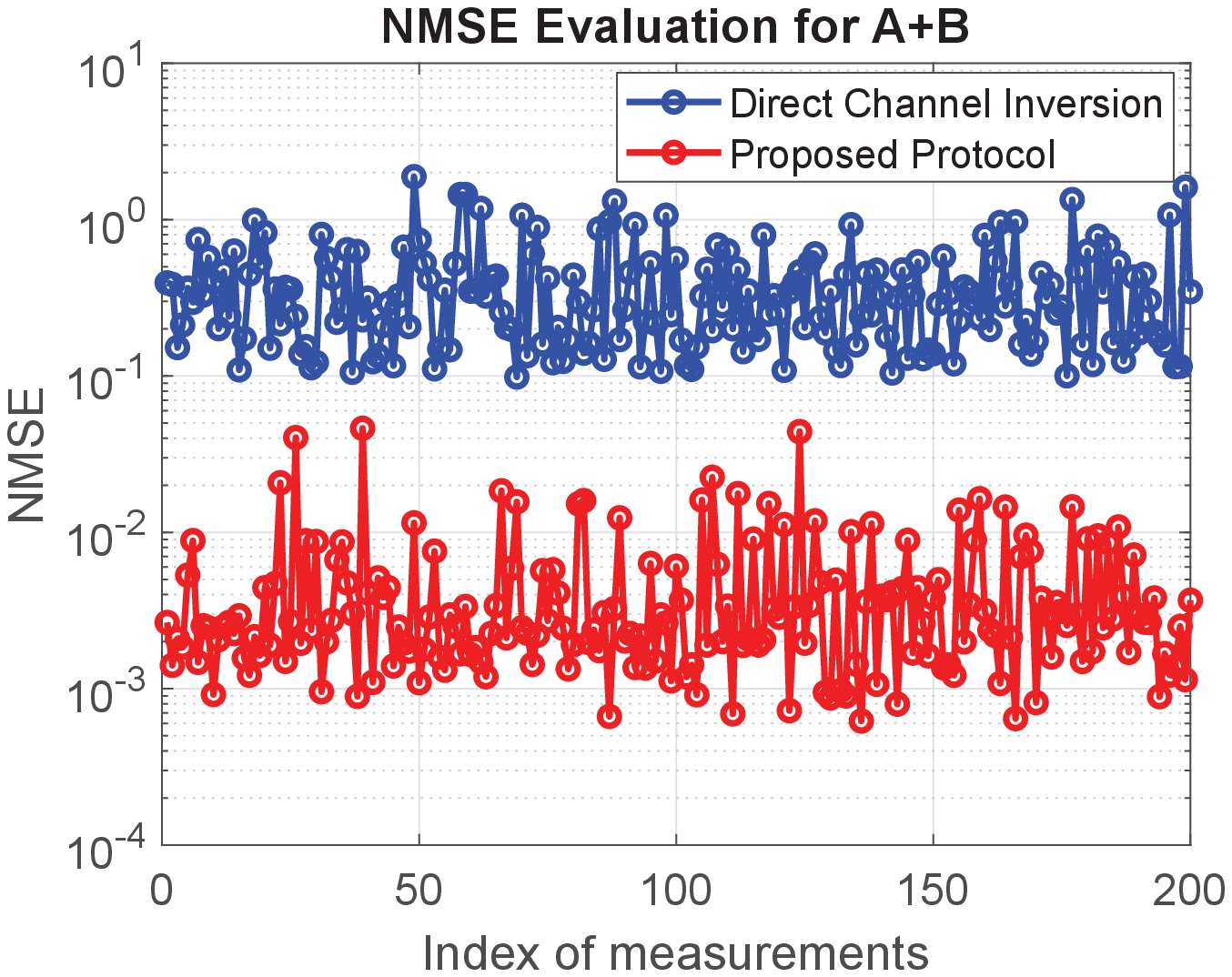}%
}
\hfil
\subfigure[]{\includegraphics[width=0.24\textwidth]{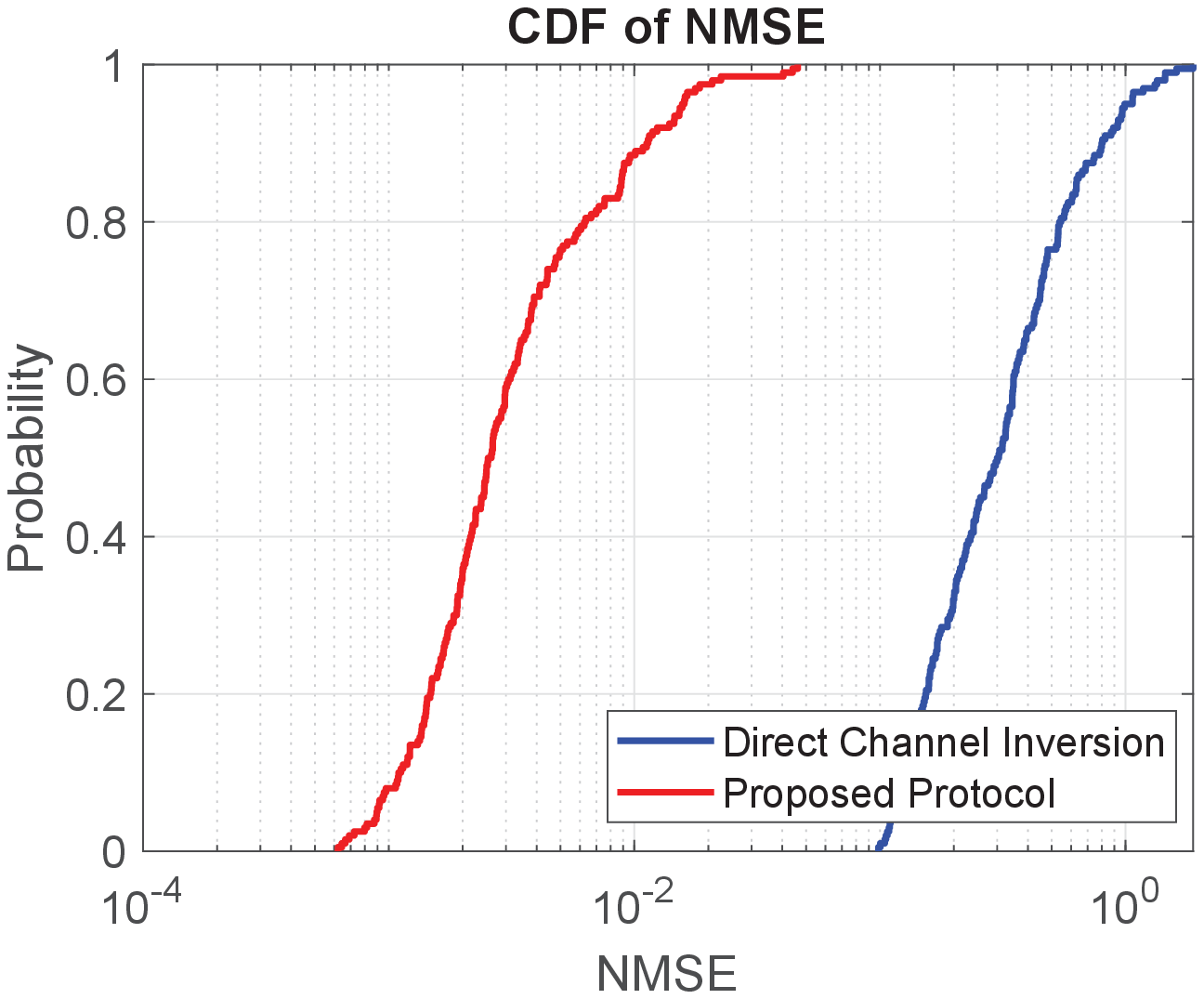}%
}
\caption{NMSE evaluation for $\mathcal{D}_{\text{OTA}}$. There are 200 measurements in this experiment and we evaluate the measruement results with/without TO and CFO compensation. (a) shows the NMSE of each measurement result, and (b) shows the cumulative probability function of the NMSE. It can be seen that 90\% of measurements can obtain an NMSE of less than 0.01 by using the proposed protocol.}
\label{fig:APB}
\end{figure}
The data pre-equalization module is for OTA aggregation purposes, which follows the method in Section \ref{sec:phase_noise_compensation}. In order to verify the performance of the proposed OTA aggregation protocol and transceiver, we perform an `A+B' test for two random number sequences by using the OTA aggregation approach. In particular, two IoT sensors transmit two independent identical distributed (i.i.d.) random number sequences $\mathbf{d}_{A}$ and $\mathbf{d}_{B}$, and the AP obtains the result $\mathbf{d}_{\text{OTA}}$ with the OTA aggregation approach. Denoting the true result of $\mathbf{d}_{A} + \mathbf{d}_B$ as $\mathbf{d}_{\text{true}}$, we evaluate the NMSE of the OTA aggregation result, which is given by
\begin{align}
    \text{NMSE}_{d} = \frac{\|\mathbf{d}_{\text{OTA}} - \mathbf{d}_{\text{true}}\|^2_2}{\|\mathbf{d}_{\text{true}}\|^2_2}.
\end{align}
As shown in Fig. \ref{fig:APB}, with the proposed protocol, the NMSE for OTA aggregation results improved significantly. The NMSE of all the results among 200 sets of measurements are less than 0.05, and this shows that 90\% of the results have an NMSE of less than 0.01.

\section{Prototype for OTA FL}
With the proposed OTA aggregation protocol and the transceiver module, we can design the prototype for OTA-FL. The proposed prototype consists of the hardware platform and software applications. The hardware platform is used for implementation of the proposed transceiver and OTA aggregation protocol, and software applications are designed to perform the FL operation. In this section, we introduce the design of the proposed prototype with one AP and two sensors to learn an NN federatively for a toy task.
\subsection{Brief Introduction to the Design}
\begin{figure} [!t]
    \centering
    \includegraphics[width=0.4\textwidth]{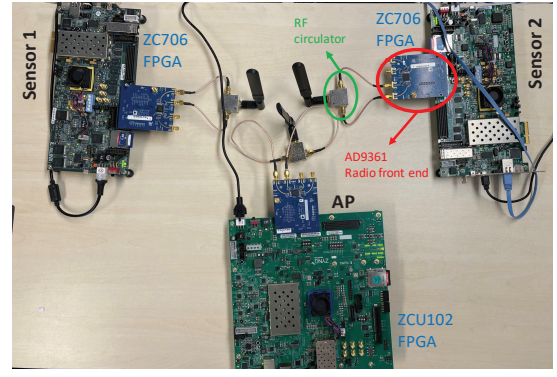}
    \caption{System Setup. There are two sensors and one AP. Each sensor uses a ZC706 FPGA for baseband signal processing and the AP uses a ZCU102 FPGA for baseband signal processing. The Radio front end is AD9361 and we use a circulator between transmitter and receiver for sharing antenna.}
    \label{fig:Setup}
\end{figure}
\begin{figure} [!t]
    \centering
    \includegraphics[width=0.4\textwidth]{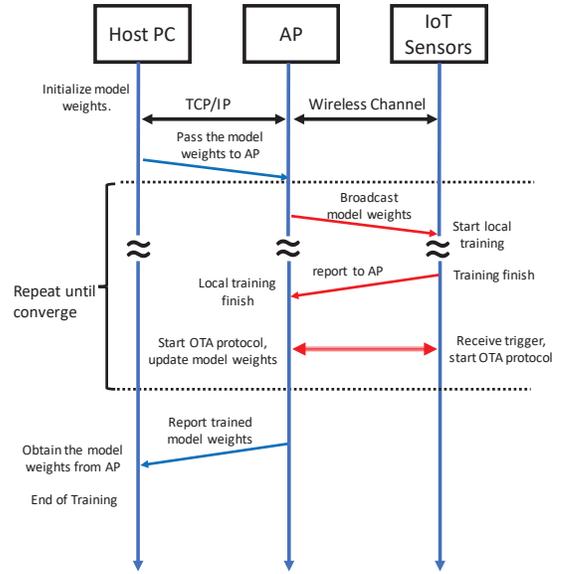}
    \caption{Application layer protocol for proposed OTA prototype. Communication between host PC and AP uses TCP/IP. }
    \label{fig:app_protocol}
\end{figure}
The structure of the hardware platform is shown in Fig. \ref{fig:app_protocol}. As shown in the figure, the platform consists of an RF front-end (RFE) and a system-on-chip (SoC) baseband processing unit. In the setup, the AD9361 is adopted as the REF, the AP uses a Xilinx-ZCU102 SoC, and IoT sensors use a Xilinx-ZC706 SoC. The system parameters are listed as follows:
\begin{itemize}
    \item The digital baseband sampling rate is 15.36 MHz;
    \item The RF baseband sampling rate is 30.72 MHz;
    \item The carrier frequency is 2.72 GHz, and the RF bandwidth is 40 MHz.
\end{itemize}\par
The software application for the FL operation follows an application layer protocol, which is provided in Fig. \ref{fig:app_protocol}. We summarize the application layer protocol as follows.
\begin{itemize}
    \item \textbf{Step 1}: The host PC initializes the model weights, and sends the weights to the AP. The host PC starts the FL training operation;
    \item \textbf{Step 2}: The AP broadcasts the model weights to the sensors;
    \item \textbf{Step 3}: The sensors perform local training, and perform OTA aggregation for the gradients;
    \item \textbf{Step 4}: The AP receives the OTA aggregation results and updates the model weights;
    \item \textbf{Step 5}: Return to \textbf{Step 2} and repeat until convergence;
    \item \textbf{Step 6}: The AP reports the final model weights to the host PC at the end of the training.
\end{itemize}

\subsection{FL problem to be learnt}
\begin{figure} [!t]
    \centering
    \includegraphics[width=0.4\textwidth]{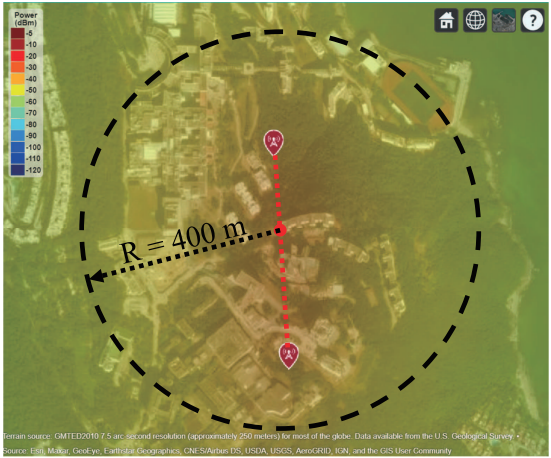}
    \caption{The RSS coverage map of the AP. The red markers denote the APs site position. We consider 2 AP sites and take 2000 RSS measurements within a circle with a 400 m radius from a center between two APs.}
    \label{fig:RSS_CovMap}
\end{figure}
In order to verify the performance of the proposed prototype, we promote an FL problem by training a fully connected NN that predicts the RSS given the GPS information (latitude and longitude) of a sensor. As shown in Fig. \ref{fig:RSS_CovMap}, we set the position of two AP sites (red markers) and generate the coverage map with the MATLAB communication toolbox. With the radio coverage map, two sensors are randomly placed within a circular area with 400 meter radius with a center between two AP sites. The sensors take 2000 RSS measurements as training data. Note that the RSS that is closer than 20 meters to the AP sites is not measured. The prediction NN is trained with an OTA aggregation approach by using the proposed prototype. The structure of the NN is shown in Table. \ref{tab:nn_sturcture}.
\begin{table}[h!]
    \caption{NN Structure for RSS Predition}
    \label{tab:nn_sturcture}
    \centering
    \begin{tabular}{c|c|c}
        \hline
        \multicolumn{2}{c|}{\textbf{Operation Layer}} & \textbf{Size of output} \\ \hline
        \multicolumn{2}{c|}{\textbf{Input GPS}} & $2$ \\ \hline
        \multirow{2}{*}{\textbf{Fc 1}} & Fully connected & $20$ \\ \cline{2-3}
        & ReLU & $20$ \\\hline
        \multirow{2}{*}{\textbf{Fc 2}} & Fully connected & $20$ \\ \cline{2-3}
        & ReLU & $20$ \\\hline
        \textbf{Fc 3} & Fully connected & $20$ \\\hline
        \multicolumn{2}{c|}{\textbf{Predicted RSS}} & $1$ \\ \hline
    \end{tabular}
\end{table}

\subsection{Finally testing}
\begin{figure}[h]
    \centering
    \includegraphics[width=3.4in]{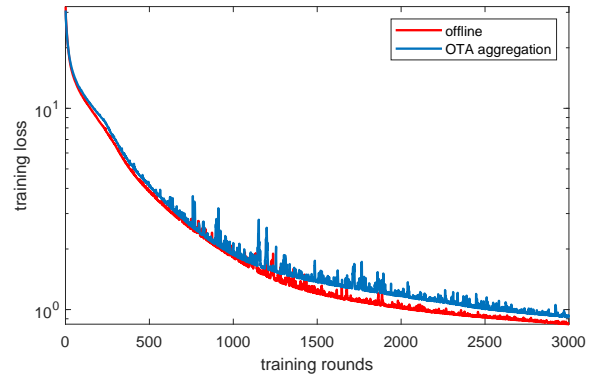}
    \caption{The training performance of the FL problem. The learning rate is set by $\eta_t = \frac{2}{2000+t}$. Each sensor stores 1000 pieces of training data. In each round, each sensor randomly picks 200 samples to calculate the local gradient.}
    \label{fig:simulation_nn}
\end{figure}

\begin{figure}[h]
    \centering
    \includegraphics[width=3in]{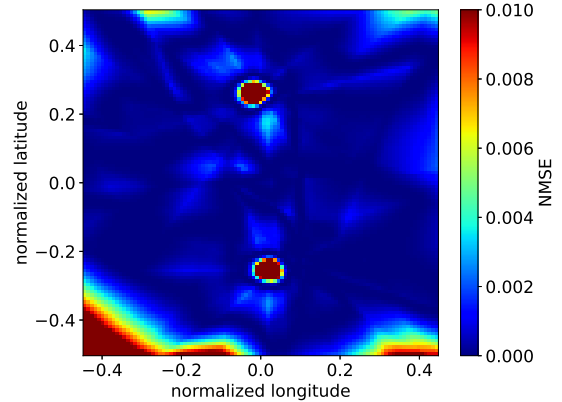}
    \caption{The heatmap of the prediction error. The NMSE of most positions is below $0.005$ with the exception of the positions that are close to AP sites due to the lack of training data.}
    \label{fig:error_heatmap}
\end{figure}

In this subsection, we use the proposed prototype to solve the promoted FL problem with the OTA method. To evaluate the performance of the proposed prototype, we show the training loss to demonstrate the convergence, and we compare the predicted RSS with the ground truth. The training result is shown in Fig. \ref{fig:simulation_nn}. We can see that our proposed OTA aggregation solution (the blue curve) achieves a similar convergence speed to the offline training (the red curve), in which the local gradients of the two sensors are aggregated noiselessly. The normalized squared predication error on the test set is shown in Fig. \ref{fig:error_heatmap}.

\section{Conclusion}
In this paper, we developed an OFDM-based over-the-air aggregation solution for a real-world federated learning task. Specifically, we analyzed the effect of frame timing offset and carrier frequency offset in the over-the-air aggregation channel, and proposed a two-stage waveform pre-equalization technique with a customized multiple access protocol to estimate and mitigate the timing offset and carrier frequency offset for the over-the-air aggregation. Based on the proposed protocol, we developed the prototype with a hardware transceiver and corresponding application software to train a deep neural network that predicts the radio signal strength with global positioning system information. To verify the performance of the proposed prototype, we performed experimental measurement and compared the learning results of over-the-air aggregation with offline learning results. From the experimental results, we can see that the proposed OFDM-based over-the-air aggregation prototype is capable for real-world federated learning tasks.
\bibliographystyle{IEEEtran}
\bibliography{IEEEabrv,OTAbib_draft2}

\end{document}